# Accessing both electrochemical SEIRA and SERS with ultrasensitive metamaterials for enhanced molecular identification

Nicolas Spiesshofer[1], Tabitha Jones[1], Sarah May Sibug Torres[1], Zoltan Sztranyovszky[2], Caleb Todd[1], Shijie Zhu[1], Yeeun Roh[1], Rakesh Arul[1], Alexander Squires[3], Ivana Qianqi Lin[4], Angela Demtriadou[2], David O Scanlon[3], Viv Lindo[5], Jeremy J Baumberg[1*]

[1] NanoPhotonics Centre, Cavendish Laboratory, Department of Physics, JJ Thompson Avenue, University of Cambridge, Cambridge, CB3 0US, United Kingdom

[2] School of Physics and Astronomy, University of Birmingham, Birmingham B15 2TT, United Kingdom

[3] School of Chemistry, University of Birmingham, Birmingham, B15 2TT, United Kingdom

[4] Hybrid Materials for Opto-Electronics Group, Department of Molecules and Materials, MESA+ Institute for Nanotechnology and Center for Brain-Inspired Nano Systems, Faculty of Science and Technology, University of Twente, 7500AE Enschede, the Netherlands

[5] Analytical Sciences, BioPharmaceuticals Development, R&D, AstraZeneca, Cambridge, United Kingdom

*email: jjb12@cam.ac.uk

## Abstract

**Surface-enhanced IR absorption (SEIRA) and surface-enhanced Raman spectroscopy (SERS) are complementary techniques that allow for ultrasensitive chemical fingerprinting. Non-invasive optical sensing would be significantly improved by a robust implementation of a reusable substrate that combines these techniques. Here, we present an electrochemically-cleanable metamaterial that enables combined real-time SEIRA and SERS in flow. This metamaterial facilitates the study of surface-adsorbed species and diffusion layers, elicits spectral shifts from changes in nanogap refractive index of 1400 nm/RIU, and delivers ultrasensitive analyte detection. Combining SERS and SEIRA clarifies molecular (electro)chemical transformations and tracks changes in selection rules and symmetry breaking at the analyte-electrode interface. This development in enhanced multimodal spectro-electrochemistry is suited for multiple domains, including understanding charge transport mechanisms and interfacial dynamics at electrodes, and is capable of real-time flow monitoring for a wide range of molecular processes.**

## Introduction

Surface-enhanced IR absorption (SEIRA) and surface-enhanced Raman spectroscopy (SERS) use plasmonic substrates that enhance incoming radiation to probe analytes at lower concentrations. They thus overcome challenges of traditional spectroscopic techniques such as strong water absorption of infrared light, limited sensitivity, and background interference. They enable continuous tracking of analyte molecular vibrational resonances and redox states, providing real-time insight into electrochemical reaction pathways, products, and interfacial reactions at electrode surfaces.[1] Using SERS and SEIRA in conjunction yields deep mechanistic insights into a wide range of molecular sensing applications[2–4] and electrochemical processes, such as the mechanism of the carbon dioxide reduction reaction ($CO_2$RR).[5,6]

Spectro-electrochemistry allows for monitoring electrochemical reactions, with electrochemical SEIRA and SERS (EC-SEIRA, EC-SERS) having applications in fields ranging from catalysis to battery research.[1,7] The adsorption of carbon monoxide,[8–10] reduction of oxygen,[11,12] carbon dioxide[5,6] and nitrogen,[13] as well as molecular catalysts[14] have been studied with IR and SEIRA spectro-electrochemistry. Additionally, measuring vibrational Stark shifts (changes in vibrational frequencies due to an externally applied potential) enables the study of local pH, the electrical double layer (EDL), and gas adsorption in EC-SERS.[8,15,16]

Combining SEIRA and SERS is crucial to obtain a complete set of molecular fingerprints as they are sensitive to different subsets of vibrational modes. However, creating electrochemical substrates combining resonant SEIRA with SERS has been challenging. Nanoantennae and nanoslit arrays can support spectro-electrochemical measurements with resonant SEIRA,[9] enabling enhancement factors up to $10^6$. However, top-down lithography methods typically fail to create reproduceable few-nanometer gaps that support high-enhancement SERS. By contrast, some SERS substrates based on chemical self-assembly or electrochemical roughening can also support non-resonant SEIRA, but only give 100-1000x signal enhancement of adsorbed species.[17] Such substrates also tend to be less robust, less reproducible, and less stable at useful potentials.[18] Self-assembled substrates combining resonant SEIRA and SERS based on metallic nanoparticles[19,17,20] open up the use of readily-fabricated metamaterial films for *in situ* low-concentration chemical sensing.

A common issue with all the above studies is degradation, which occurs due to irreversible binding of analytes (such as gases and biomolecules) to the SERS/SEIRA substrate, affecting any subsequent measurements with the same substrate. 'ReSERS'[21] is a recently developed method that uses electrochemical oxidation of gold nanoparticles to repeatedly strip off analytes, maintaining low variability (<5%) across samples over many use cycles. By contrast, *in situ* electrochemical regeneration for SEIRA has not yet been achieved, due to the lack of suitable substrates. Initial work to develop reusable SEIRA sensors, includes the use of regenerated liquid gallium[22] and protective polymer layers[23] for targeted applications. However, this remains an underdeveloped area within the field.

Here, we report an EC-SEIRA/SERS substrate that uses multilayer aggregates (MLaggs) of precision-spaced AuNPs supported on graphene-capped ZnSe for electrochemical measurements in solution at trace concentrations. We show that this substrate is reusable via electrochemical oxidation/reduction thereby removing all analytes. We test this substrate by monitoring the model redox reaction of ferrocyanide detecting both SEIRA and SERS. This ferrocyanide/ferricyanide redox couple has distinct IR and Raman peaks, which change depending on the oxidation state of the analyte.[18,24] We find that monitoring the redox couple via both SEIRA and SERS yields insights into the diffusion-limited behavior within the tortuous metamaterial and highlights symmetry breaking of the molecule at electrodes. The resulting emergence of simultaneously IR- and Raman-active species bound at metal surfaces has implications for plasmon-driven chemistry and catalysis, for example in improving hot electron transfer.[25]

## Results and Discussion

### MLaggs on graphene as reusable EC-SEIRA sensors

The assembly of gold nanoparticles into close-packed aggregate films (MLaggs) as well as their stacking to obtain tuneable resonances has been shown in previous work.[17,19] These precision-spaced 100 nm AuNPs separated by 0.9±0.05 nm gaps defined by cucurbit[5]uril (CB[5]) spacer molecules, form a random close-packed metamaterial (**Figure 1a-c**). These MLaggs give strong resonances with intense optical hotspots inside each nanogap. Sequential deposition of up to 10 stacked MLaggs (10MLs) red-shifts the resonance into the molecular fingerprint mid-IR spectral region, and can be further red-shifted using larger 250 nm diameter AuNPs (**Figure 1d**). The SEIRA vibrational resonances of CB[5] exhibit clear Fano lineshapes which tune in amplitude and phase with the number of MLagg layers and AuNP size, in contrast to previous non-resonant EC-SEIRA substrates.[8,11,14] Their strong IR enhancement allows detection of a range of small molecules at low concentration in solution, such as pharmaceutical compounds (amantadine) or DNA bases (SI Figure S1).

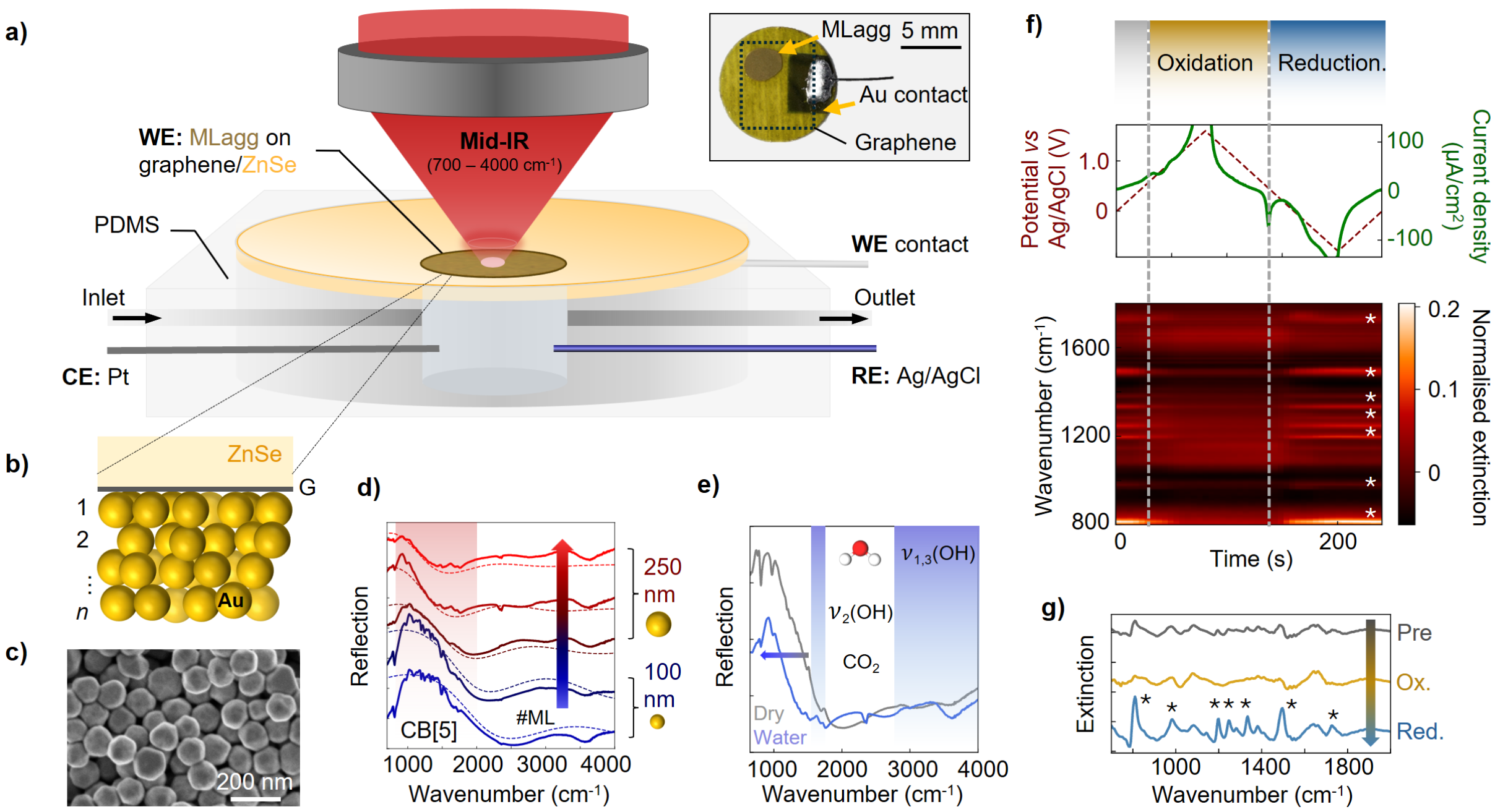

**Figure 1. a**) EC-SEIRA flow cell, showing flow setup with inlet, outlet, and sample well. Labelled working electrode (WE), reference electrode (RE) and counter-electrode (CE), with inset showing image of device, graphene region dashed. **b**) Layers of MLagg below graphene (G) and ZnSe. **c**) SEM of MLagg on graphene/ZnSe. **d**) Tuning of mid-IR resonance of MLaggs in water for different layer number and nanoparticle size, with transfer matrix fits (dashed). **e**) Dry *vs* aqueous mid-IR reflectivity with water absorption bands highlighted. Arrow shows red-shift of resonance. **f**) SEIRA spectro-electrochemistry showing regeneration of substrate and cyclic voltammetry. Oxidation and reduction sections labelled. **g**) *In situ* electrochemical cleaning of MLagg (Pre=before cleaning), showing stronger molecular vibrational resonances of CB[5] after ReSERS (Red.). In (f),(g), asterisks (*) label CB[5] vibrations.

To introduce electrochemical control with back-side NIR/MIR access and thus develop EC-SEIRA and EC-SERS, we deposit AuNP MLaggs on the IR-transparent conductive monolayer graphene (2D/G ratio > 2, SI Figure S2,3) which is coated onto ZnSe and connected via Au contacts and Pt wiring (**Figure 1a**). The light absorption of this conducting support is identical to ZnSe since graphene absorbs minimally in the visible and mid-IR spectral region.[18] The use of ZnSe is advantageous as its transparency from ~600 nm to the far-IR avoids masking molecular vibrations that are obscured with other substrates such as $SiO_2$. Crucially it also allows NIR light to penetrate, thus accessing the MLagg plasmonic resonances that give intense SERS. As shown below, the MLaggs electrically contact the graphene layer, providing an EC-metamaterial on a fully transparent substrate. These devices are used in a compact microfluidic cell, which integrates flow channels and electrochemical electrodes and fits underneath an IR/Raman microscope (**Figure 1a**).

The deposition of MLaggs on graphene (G/ZnSe) leads to resonances in the mid-IR, which for optimal sensing should be tuned to the desired MIR vibrations. We have shown that since the wavelength (5-10 µm) is much larger than the NP size and spacing, the MIR optical field distribution can be understood by treating the MLagg metamaterial as a high-refractive-index effective medium[26] and applying transfer matrix methods (adapted from[17]). This gives distinct peaks in reflectance (**Figure 1d**) and absorption. These match the reflectance spectra closely and can be used to predict the electric field envelope in the metamaterial, which has a maximum at the ZnSe/metamaterial interface and an antinode at the metamaterial-air/water interface (SI Figure S4). Conversely, the field-envelope in the near-infrared ($\lambda$ = 785 nm) exponentially decays over the depth of the metamaterial, with most of the field only present in the top MLagg layer (SI Figure S5). Within this field envelope, the light is further highly concentrated within the plasmonic hotspots, with the location of hotspots probed depending on the resonance tuning.

Crucially, we now find that the water absorption bands do not obscure the CB[5] vibrational resonances or those of analytes, which is essential for sensing in aqueous conditions. Specifically, the O-H bending mode ~1600 $cm^{-1}$ is only double the 1750 $cm^{-1}$ carbonyl CB[5] absorption (**Figure 1e**), showing that these substrates are suitable for IR flow sensing, which is replicated with other solvents (SI Figure S6).

The electrochemical properties of the G/ZnSe support (without MLaggs) show excellent conductivity and stability over a wide voltage window (at least -0.8 V $<V<$ 1.5 V *vs* Ag/AgCl, SI Figure S7, larger anodic potentials possible[27]). To demonstrate the reusability of the substrate, we remove surface-bound species by sweeping the applied potential over the gold oxidation potential, followed by a gap 'rescaffolding' step via gold reduction in the presence of CB[5]. Cyclic voltammetry (CV) in 1 M pH 7 potassium phosphate buffer (KPB) and 1 mM CB[5] with 5 $mVs^{-1}$ scan rate shows characteristic anodic and cathodic current peaks corresponding to the oxidation and reduction of Au,[21] indicating that the MLagg surfaces are oxidized and reduced (**Figure 1f**). Oxidizing potentials (>+1.0 V) lead to a clear decrease in the amplitude of all CB[5] vibrational resonances (**Figure 1g**), indicating desorption of surface-bound species as the MLagg hotspot becomes plugged by gold oxide.[21] Reducing potentials (<+0.3 V) in CB[5] solution have the opposite effect, with all CB[5] vibrational resonances restored. By oxidizing and reducing the MLagg after initial fabrication, it is completely pre-cleaned and thus primed for sensing (**Figure 1g**). After this pre-cleaning, the SEIRA signal of CB[5] has now doubled due to removal of other surface-bound species (such as citrate surfactant), enhancing Au surface coverage by CB[5]. Adsorbed analytes can also be subsequently removed from the nanogaps, regenerating the surface to its original state and allowing continuous flow sensing (as shown for ferricyanide, SI Figure S8).

**Electrochemical regeneration effects on molecular vibrations and MIR resonances**

To assess the stability and reusability of the substrate over time, we perform multiple oxidation and reduction sweeps, demonstrating the practical use for continuous monitoring. We monotonically cycle the potential to oxidize and rescaffold the MLagg five times in 1 M, pH 7 KPB and 1 mM CB[5] at 5 mVs$^{-1}$ with respect to Ag/AgCl reference-and Pt wire counter-electrode, while observing changes in molecular vibrations coupled to the MIR reflectance mode. We find that cycling between Au oxidation and reduction does not adversely affect the MLagg resonance or the electrode surface, with the metamaterial resonance staying stable over multiple cycles and repeatable molecular vibrational signals (**Figure 2a,b**).

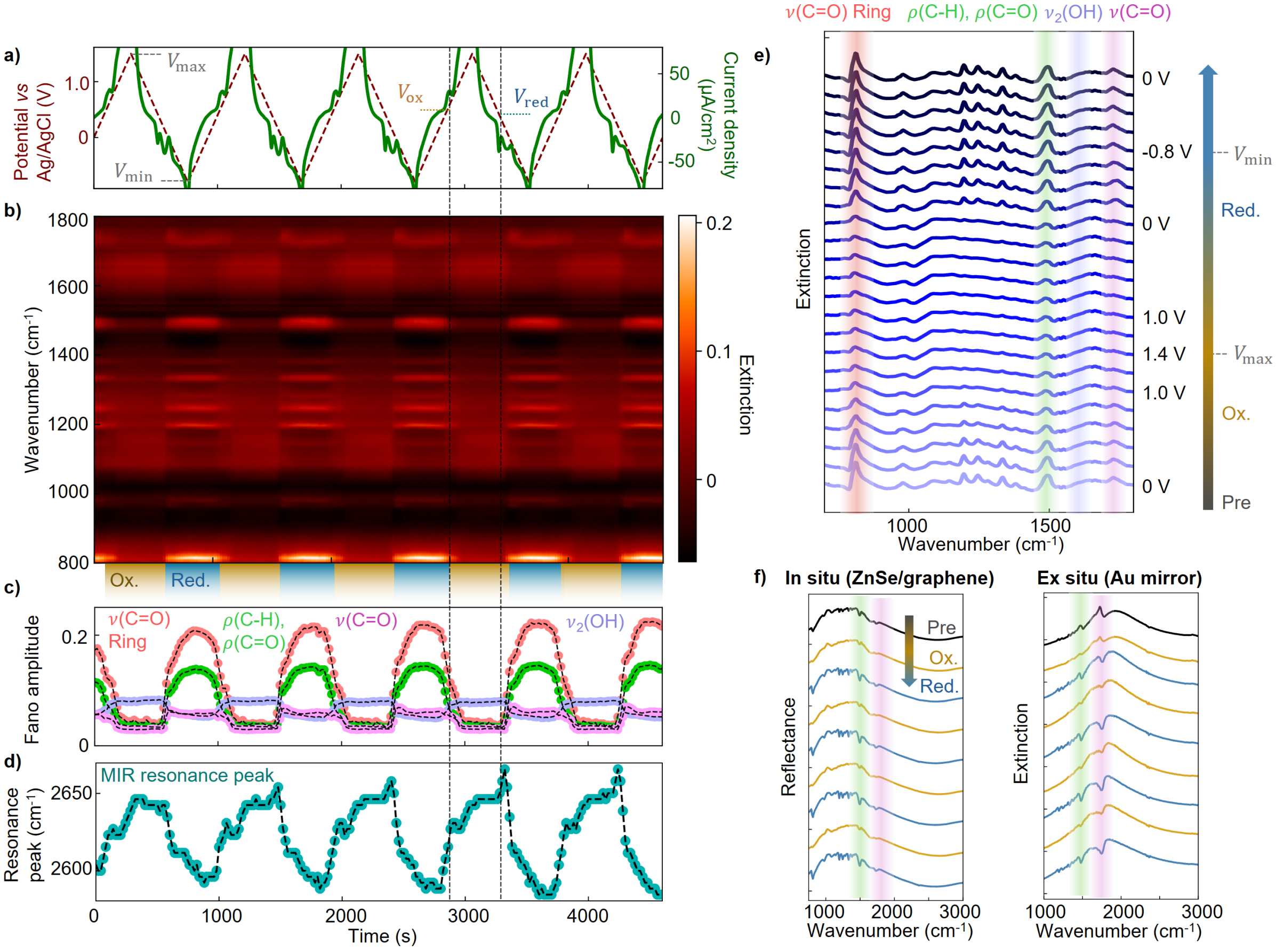

**Figure 2. a**) Cyclic voltammetry over five cycles. **b**) SEIRA spectro-electrochemistry showing regeneration of substrate over five cycles of oxidation (yellow shaded below) and reduction (blue shaded). **c**) Fano-fitted CB[5] vibrational resonances (red, green, pink) and water absorption (purple). **d**) Change in peak position of MLagg mid-IR resonance during cycling. **e**) SEIRA spectra over a single regeneration cycle, vibrational resonances as indicated. **f**) Comparison of *in situ* and *ex situ* electrochemical regeneration. In (c),(e), $\nu$ represents IR stretching and $\rho$ rocking vibrations, assignments from [17].

We find that all CB[5] vibrational resonance changes correlate with the CV when the MLagg is oxidized (**Figure 2c,e**), nearly fully disappearing (reduced by >80%). Moreover, rapid recovery of gap-bound species is seen even in these layered closely-packed nanostructures when sweeping the voltage between oxidation and reduction at <10 mVs$^{-1}$ scan rates,[28] suggesting that diffusion into the nanogap hotspots is maintained.

A repeatable blue-shift of the MLagg MIR resonance is observed as the sample is oxidized *in situ* (**Figure 2d,f**), however when analyzed *ex situ* (oxidized *in situ*, dried, then analyzed) the resonance is seen to red-shift, which suggests an undesirable structural reconfiguration as the sample is dried (Figure 2f, SI Figure S9,10). Moreover for the latter, a gradual decrease in CB[5] SEIRA strength, changes in Fano lineshape, and broadening of the MIR resonance is seen, indicating progressive sintering.[29] No such effect is visible when cycling *in situ*, showing this is preferable.

**MIR resonance shift tracks reconfiguration of Au surfaces**

As the MLagg is oxidized and reduced, its MIR resonance changes in a characteristic and repeatable manner, indicating subtle changes to the nanoparticle surface (**Figure 2d, 3a,b**). We observe a blue-shift as the sample is oxidized, which is reversed when the MLagg is reduced (SI Figure S11).

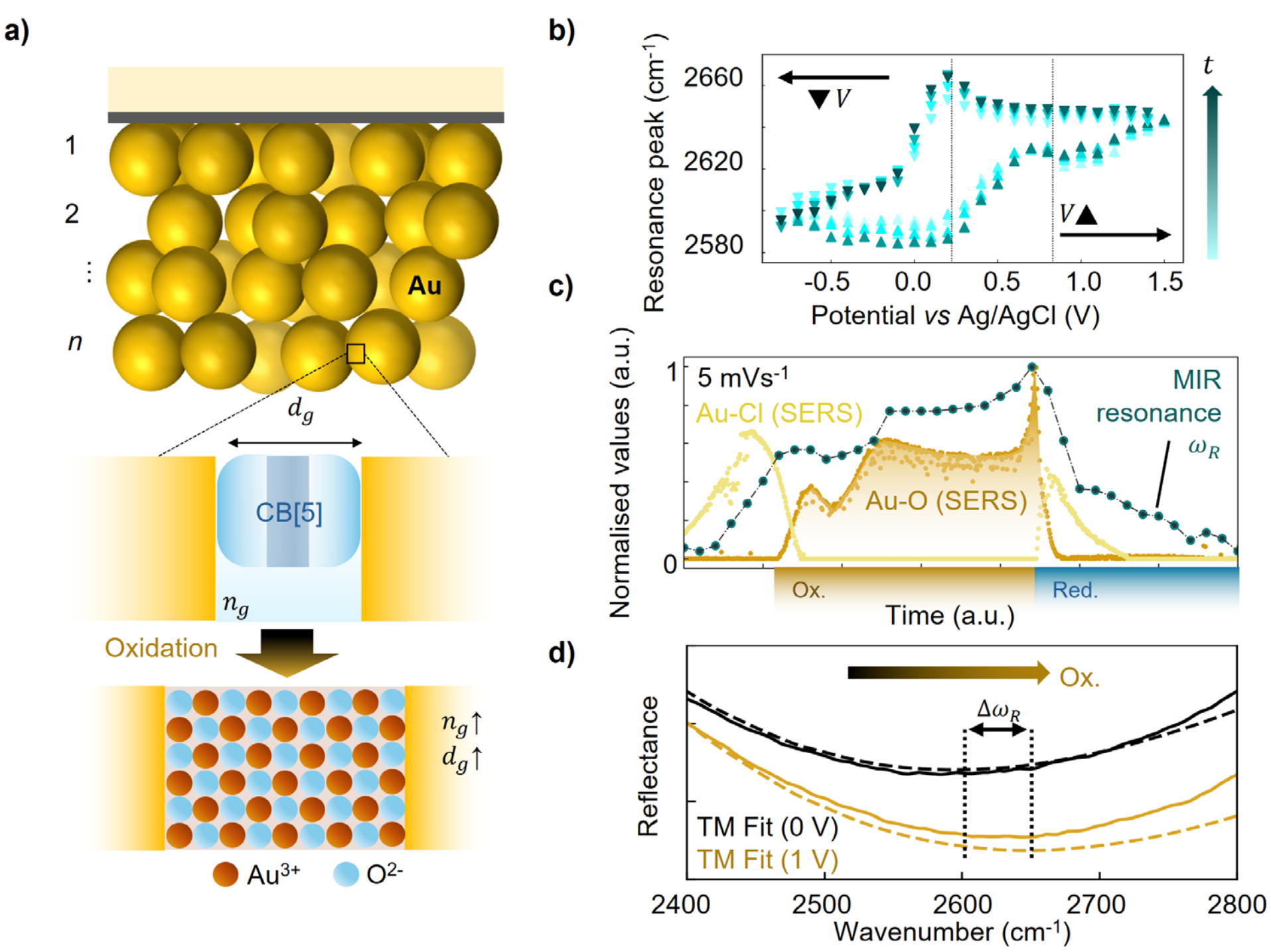

**Figure 3. a**) Shifts in MIR resonance peak position over successive cycles *vs* applied potential. **b**) Schematic changes in gap size and refractive index as CB[5] is replaced by Au(III) oxide ($Au_2O_3$) during oxidation. **c**) Comparison of SERS Au-O and Au-Cl signals (lines, equivalent experimental conditions, 1-2ML MLagg on FTO, corrected for different open circuit potential on FTO and graphene) through an ox/red. cycle, as well as SEIRA resonance peak position (blue points), which is dependent on the gap refractive index and changes with the presence Au-O and Au-Cl. **d**) Combined transfer matrix (TM) and dielectric oscillator model (including Au-O), before and after oxidation, showing shift $\Delta\omega_R$ in reflectance dip.

We observe a strong correlation between these shifts in the MIR resonance peak frequency (blue, **Figure 3c**) and the strength of the $Au_2O_3$ peak area in SERS (brown).[21] This provides evidence that the MIR spectral shifts track nanoscale changes of the gap contents (as illustrated in **Figure 3a**). Gold chloride (Au-Cl) further influences the resonance position as its coating of the nanoparticle surface at positive potentials[28] leads to changes in the surrounding dielectric environment, giving also strong SERS signals.

We attribute the increase in resonance frequency during oxidation to a combination of: i) an increase in the size of the gap between Au(0) surfaces of the nanoparticles (**Figure 3a**),[19] ii) an increase in the metamaterial effective refractive index as gold oxide fills the gaps,[26] and iii) the introduction of multiple strong broad infrared-active $Au_2O_3$ absorption lines[30,28] (DFT simulation SI Figure S12) resulting from lattice vibrations at ~600 $cm^{-1}$ whose tail affects the MIR resonance position (SI Figure S13). By fitting a dielectric oscillator model to the gold oxide IR peak (see discussion SI Note 1), we find that indeed increasing lower frequency $Au_2O_3$ absorption shifts the MIR resonance to higher frequency (**Figure 3d**).

We obtain similar effects by fitting the metamaterial effective refractive index and gap thickness to match the final shift in resonance. While an increase in refractive index will red-shift the resonance, an increase in gap size due to the conversion of several gold monolayers to Au(III) oxide leads to a lower effective refractive index, blue-shifting the resonance.[26] Previous estimates show that the gap increases from 0.9 to 1.8 nm due to the formation of two to three atomic layers of Au(III) oxide (previously confirmed by XPS[28]). The change in the effective metamaterial index $n_{\text{eff}}$ due to this change in gap refractive index $n_g$ has been previously[26] modelled by:

$$n_{\text{eff}}^2 = n_g^2 \frac{1+2f}{1-f} \tag{1}$$

The change in fill fraction $f$ from a gap size of 0.9 nm to 1.8 nm in a hexagonal close-packed lattice is from 0.72 to 0.68. Based on this change and the measured ~3% change in resonance frequency, we estimate an increase in the gap refractive index by ~8% based on an initial estimate of $n_{\text{eff}} \approx 3.1$ and $n_g \approx 1.1$ from previous fitting[17,26] (see discussion SI Note 1). The sensitivity to tracking changes in the nanogap refractive index thus amounts to ~1400 nm/RIU, using the underlying assumptions mentioned above. This matches advanced fiber-based SPR systems (>1000 nm/RIU in the visible) and multilayer grating based SPR sensors (~300-2500 nm/RIU in the visible),[31] and is larger than qBIC metasurfaces.[32] Hence, these colloidal metamaterials track sensitively the sub-nanometer changes in surface composition.

**Complementary molecular sensing with SERS and SEIRA at nanoscale interfaces**

These EC-devices are versatile sensors to access chemical mechanistic insights involving interfacial reaction dynamics. By using MLaggs as the working electrode, molecules can interact with the gold surface in many ways, including adsorption and electron transfer. Here we use complementary SEIRA and SERS on the same device (**Figure 4a**) to sense the coordination compound ferrocyanide, $[Fe(II)(CN)_6]^{4-}$, down to concentrations <1 mM in 500 mM potassium phosphate buffer (KPB) with high signal-to-noise for SEIRA and SERS (SI Figure S14). We probe electron transfer as it switches its redox state to become ferricyanide, $[Fe(III)(CN)_6]^{3-}$ (**Figure 4a**).

Cycling the redox couple from -0.7 to +0.6 V at 2 $mVs^{-1}$ (**Figure 4b**) shows both changes in oxidation state and surface interactions whilst keeping the potential below that for gold oxidation. We confirm that the cyclic voltammetry is similar to that on planar FTO, showing that the redox couple behaves in an expected manner on these metamaterial sensors (SI Figure S15).

Through cyclic voltammetry combined with SEIRA and SERS measurements, we observe subtle changes in the interaction of the various ferricyanide species on the electrode surface. We see strong $[Fe(II)(CN)_6]^{4-}$ signals in both SEIRA and SERS despite the additional presence of CB[5] and counter-cations. Firstly, we observe a $[Fe(II)(CN)_6]^{4-}$ $\nu$(C≡N) mode at ~2065 $cm^{-1}$ that is SEIRA-active but SERS-inactive. The second $[Fe(II)(CN)_6]^{4-}$ $\nu$(C≡N) mode at 2115 $cm^{-1}$ is both SEIRA- and SERS-active. A third $[Fe(III)(CN)_6]^{3-}$ $\nu$(C≡N) mode at ~2170 $cm^{-1}$ is stronger in SERS than SEIRA. We also track the decrease in characteristic CB[5] lines after immersion in $[Fe(II)(CN)_6]^{4-}$ solution, resulting from adsorption of $[Fe(II)(CN)_6]^{4-}$ and desorption of CB[5] (SI Figure S14). Further data collected in the SERS low-wavenumber spectral region, reveals peaks for Fe-C and Fe-C≡N-Au modes (SI Figure S16).

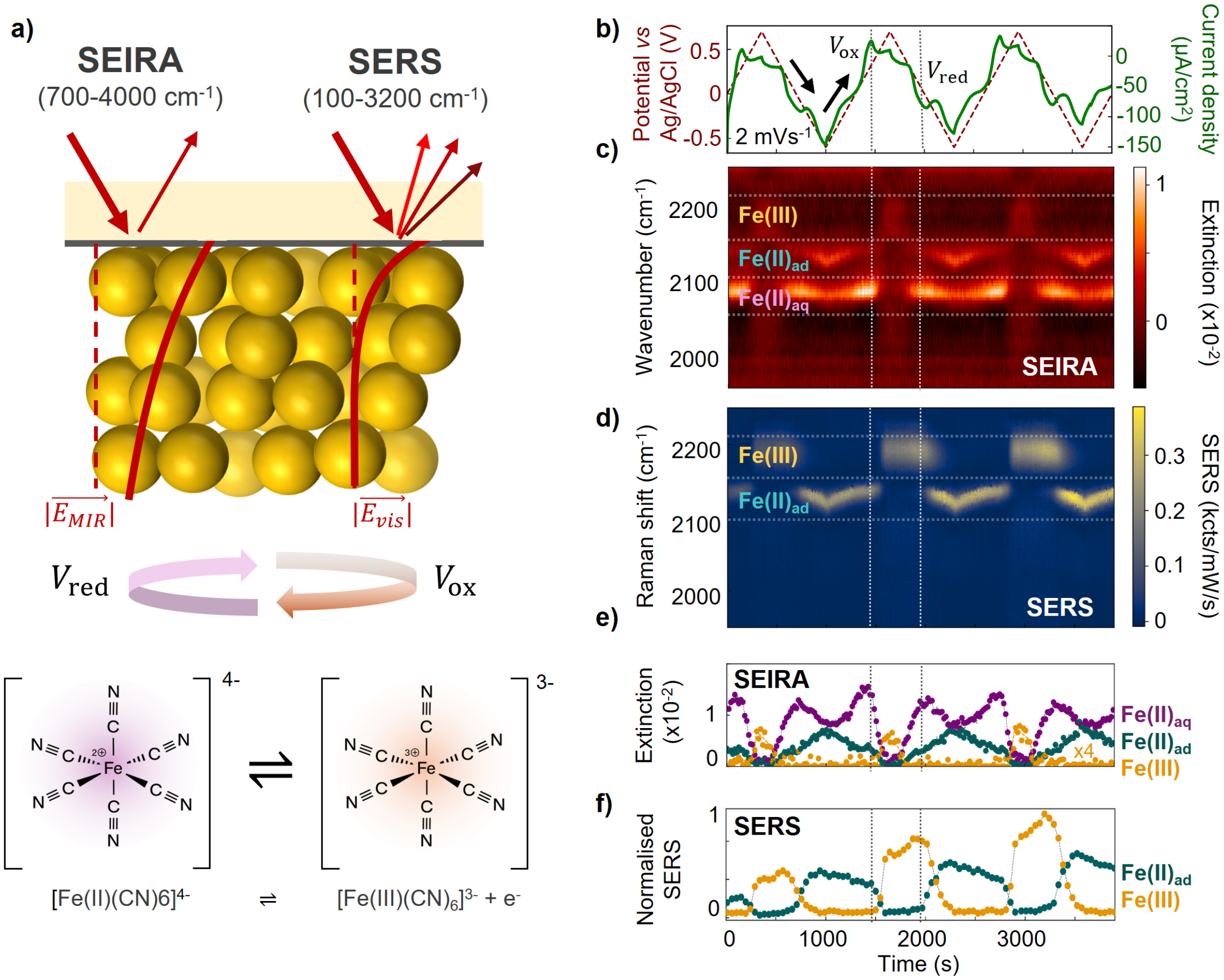

**Figure 4. a**) Schematic of metamaterial modes for SEIRA and SERS, with field distributions at the MIR and visible wavelengths. **b**) Cyclic voltammetry of ferro/ferricyanide $[Fe(II)(CN)_6]^{4-}$/$[Fe(III)(CN)_6]^{3-}$ redox couple (5 mM in 500 mM KPB), over three cycles; dashed lines indicate position of ferricyanide oxidation and reduction peaks. **c-d**) Spectro-electrochemistry of ferrocyanide couple with (c) SEIRA extinction and (d) SERS. **e-f**) Extracted fits of (e) SEIRA and (f) normalized SERS peak areas.

By comparing these highly sensitive SEIRA measurements to SERS on MLaggs, we obtain detailed complementary information about the redox couple. Key puzzles in this data are (i) a Stark-shifting mode appearing in both SERS and SEIRA at 2115 $cm^{-1}$; (ii) the different potential dependence of the same Fe(III) 2170 $cm^{-1}$ mode in SERS and SEIRA; and (iii) the different potential dependence of different infrared-active Fe(II) modes at 2065 and 2115 $cm^{-1}$.

## Mechanism of ferrocyanide couple in nanogaps

We perform DFT of likely configurations (**Figure 5a-b**). From the experiment, we find one vibrational mode (2115 $cm^{-1}$) that is common to both SEIRA and SERS. This contrasts with the calculated normal modes for isolated ferrocyanide in the mid-IR ($F_{1u}$) and Raman ($A_{1g}$, $E_g$) with complementary symmetries[33] (**Figure 5a-b,** SI Figure S17-19). This peak is blue-shifted compared to the calculated normal modes, which is indicative of close proximity of the cyanide functional groups to the gold NP surface (**Figure 5c,** labelled A), resulting in a bridged C≡N-Au mode (that we thus term $Fe(II)_{ad}$).[34] From polarized DFT (vibrational modes aligned with the field normal to the metal surface are selected), the $\nu$(C≡N) mode indeed shifts to higher wavenumbers due to σ-donation to the N-Au (**Figure 5d-e**). Conversely, this leads to a red-shift in the Fe-C low-frequency mode, which we also observe both in our data and DFT (SI Figure S16,17).[35] Low-wavenumber SERS further reveals the emergence of multiple bands linked to ferrocyanide-metal bridging (SI Figure S16,18).

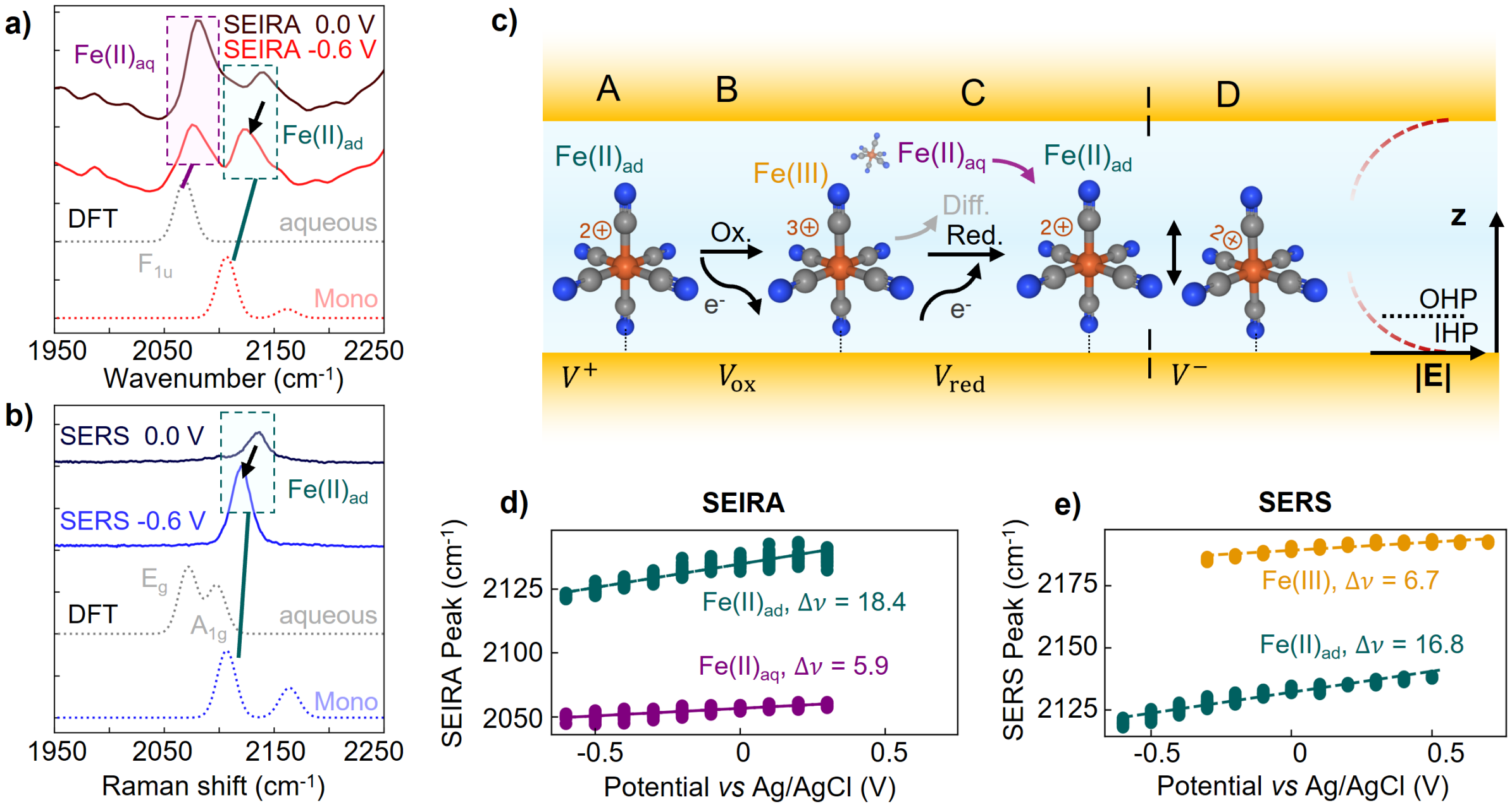

**Figure 5. a-b**) Change in (a) SEIRA and (b) SERS at different potentials, with DFT simulated peaks for solution-phase (aqueous) and polarized DFT aligned along normal to metal surface for Fe(II) adsorbed species, bound by one N (monopod). **c**) Schematic of ferro-/ferricyanide species on the charged gold nanoparticle surface, with alphabetic labels to guide the reader; OHP, outer Helmholtz plane; IHP, inner Helmholt plane. **d-e**) Extracted peak shifts for (d) SEIRA and (e) SERS.

A likely interpretation for the emergence of a peak that is both IR- and Raman-active, lies in the broken symmetry of ferrocyanide close to metallic surfaces. Related observations have been seen in the SERS and SEIRA of CB[5] in MLaggs[17] and for other coordination compounds such as tungsten hexacarbonyl.[36] In a nanogap 0.9 nm wide, ferrocyanide is most likely present as a monolayer between the two AuNP surfaces (as its length is 0.62 nm [33]). At these distances, we expect distortion of the octagonal symmetry of ferricyanide,[37] as previously demonstrated in SERS[35] and IR[24] alone using roughened electrodes, which lack the precise faceting of AuNPs. The bridging interaction of the AuNP surface with ferrocyanide leads to different orientational configurations (some explored in SI Figure S17,18). The monopod configuration (one nitrogen atom bound to the gold surface) shows the best agreement between DFT and SEIRA/SERS (**Figure 5a-c**, SI Figure S17). We here expect the emergence of one common Raman- and IR-active mode, A1, with additional Raman- and IR-active modes (as shown in the data and polarized DFT**, Figure 5d-e**). These other possible modes do not appear in SEIRA/SERS due to surface-selection rules (the optical fields are always predominantly normal to the metal).[36]

Comparing the vibrational peak shifts for the different species (**Figure 5d,e**), we note that the A1 mode of $Fe(II)_{ad}$ exhibits a significant nonlinear shift, whereas the $Fe(II)_{aq}$ seen only in SEIRA does not. The main origin is a change in the binding strength of the Au-N coordination bond, which modulates the peak frequency (as sketched in **Figure 5c**, labelled D). The counter-ions in the EDL may further screen the cyanide bond adjacent to the gold surface, changing the polarizability of ferrocyanide.

We assign the additional lower-wavenumber peak (2065 $cm^{-1}$) in SEIRA to the solution-phase $Fe(II)_{aq}$ species, which has no broken symmetry and thus appears only in SEIRA. Further evidence lies in its smaller Stark shift, which only occurs close to the electrode surface. No SERS-active solution-phase species are observed, likely due to the stronger concentration of field in the nanogap for SERS compared to SEIRA, which probes further out of the nanogap. The correlated changes in $Fe(II)_{ad}$ and $Fe(II)_{aq}$ across the potential cycle can be understood by considering the diffusion layer within the MLagg. For the MLagg at more negative potentials, more Fe(II) adsorbs at the gold (**Figure 5c**, labelled C), leading to an increase in the local $Fe(II)_{ad}$ concentration and corresponding decrease in the $Fe(II)_{aq}$ signal (SI Figure S20). This forms a diffusion layer whose dynamics are hence monitored with SEIRA.

As $Fe(II)_{ad}$ is rapidly oxidized, Fe(III) builds up and Fe(II) is depleted (**Figure 4c-e, 5c**, labelled B), indicated by the 2170 $cm^{-1}$ mode. Our data suggests that again surface-bound species are present, this time for Fe(III), with symmetry breaking allowing both Raman and IR absorption (SI Figure S18). Surprisingly, as the applied potential reverses, SERS and SEIRA show the Fe(III) disappearing at different potentials, despite their similar peak positions. This implies that during this reduction to Fe(II), Fe(III) species are located in different regions, with some remaining in the diffusion layer and getting reduced at lower potentials (**Figure 4b**). We propose that the different potential thresholds seen in SERS and SEIRA stem from the distinct optical field envelopes of SERS and SEIRA (**Figure 4a**, SI Figure S4,5), with SEIRA probing much further into the MLagg where the potential can be lower than at the graphene contact due to the potential drop between individual gold nanoparticles.

Despite the well-studied nature of this ferrocyanide redox couple, we show here that combining surface-enhanced IR and Raman spectroscopies reveals more than previously known about symmetry breaking of molecules at electrode surfaces. Detailed spectro-electrochemistry results give new data to test against a variety of theoretical models, that moreover should combine both surface binding, ionic details of the EDL, as well as dynamics of the diffusion layer. It shows how the complementarity of SEIRA and SERS can be exploited in a multiplexed platform for future studies of different molecules, such as in novel battery electrolytes.

## Conclusion

This study presents an architecture to perform real-time combined SEIRA and SERS measurements on a plasmonically-active metamaterials device. It presents the first electrochemical regeneration method for SEIRA substrates, evidenced by the replacement of the molecular linker molecule by an atomically thin layer of gold oxide, whose generation can be monitored through changes in the nanogap effective refractive index. These results further demonstrate the potential of using complementary electrochemistry, SEIRA, and SERS measurements on the same flow device, allowing for easier and more reliable measurement and analysis of redox reactions. We emphasize that tracking such symmetry-breaking behavior on metallic electrode surfaces is of great interest in plasmonic catalysis and plasmon-driven chemistry, as well as real-time monitoring the pharmaceutical production of bioproducts.[38,39]

## METHODS:

**Graphene substrate fabrication**. The ZnSe substrate was oxygen plasma cleaned for 5 seconds (oxygen mass flow of 30 sccm, 90% RF power) using a plasma etcher (Diener electronic GmbH & Co. KG) to remove surface-bound organics and improve its hydrophilicity. The monolayer Trivial Transfer® CVD graphene (ACS Materials) was then transferred onto the clean ZnSe substrate, dried for 2 h and then baked for 20 min at 100°C. Subsequently, the PMMA adhesion layer was dissolved by immersion in acetone for 24 h, rinsed, 15 min, rinsed, 5 min. Graphene quality and the removal of PMMA residues was checked with Raman spectroscopy, using a Renishaw InViva Raman confocal microscope, using a 100x objective and 1200 lines $mm^{-1}$ grating, and a 532 nm excitation laser. A gold contact was thermally evaporated (2 nm Cr, 50 nm Au) using a shadow mask with a deposition rate <0.5 A/s. The contact was then attached to a Pt wire via Field's metal heated to 65°C, forming a stable connection. Finally, the contact was insulated with nail varnish, leaving only the pristine graphene exposed.

**Multilayer aggregate formation.** 100 and 250 nm gold (BBI Solutions) nanoparticles were mixed in an Eppendorf tube to make up a total of 500 µL of nanoparticles. 500 µL of chloroform ($CHCl_3$) was then added to another Eppendorf, followed by 500 µL of the nanoparticle mixture. 100 µL of 1 mM cucurbit[5]uril (CB[5]) solution was then added and shaken for ~1 min to initiate aggregation. The mixture was left to settle for the immiscible $CHCl_3$ and aqueous phases to separate and the aggregated AuNPs to move to the phase interfaces (chloroform-aqueous and aqueous-air). The aqueous phase was washed with three 300 µL aliquots of DI water to dilute the citrate salts and other supernatants, then concentrated by careful removal of the aqueous phase to form a ~5 µL aggregate droplet floating on the $CHCl_3$. The droplet was deposited onto the graphene substrate and left to dry. Once dried, the resulting AuNP multilayer aggregate was rinsed with DI water and dried with $N_2$. This was repeated another 4 times, depositing the new aggregate droplet in the same position each time for a total of 5 depositions to give a sample with multiple layers of nanoparticles.

**Electrochemical measurements**: The sample was immersed in an electrochemical flow cell. The cell is fabricated out of 10:1 PDMS:cross-linker, and its geometry adjusted via biopsy punches to include a well with a diameter of 6 mm, inlet and outlet, as well as inserts for electrodes. The three-electrode amperometric system included a Pt counter electrode and Ag/AgCl reference electrode. The sample was used as the working electrode. The current density was obtained by dividing the current by the area of the exposed working electrode (~0.28 $cm^2$). Measurements were performed using a portable potentiostat (CompactStat) from Ivium Technologies.

**Plasma cleaning:** The MLaggs are oxygen plasma cleaned for 45 minutes (oxygen mass flow of 30 sccm, 90% RF power) using a plasma etcher (Diener electronic GmbH & Co. KG) to remove CB[5], citrate and other supernatants from the AuNP surfaces (verified using SERS). To re-introduce a scaffolding ligand, the MLaggs are immersed in equal volumes of 1 mM CB[5] or CB[7] solution and 1 M HCl for 5 minutes. In both cases, the MLaggs are rinsed with DI water and dried with $N_2$.

**Infrared microscopy:** All samples are analyzed in a Shimadzu AIM-9000 infrared microscope, linked to a Shimadzu IR-Tracer 100. For image acquisition, a spectral resolution of 8 $cm^{-1}$, and aperture size of 200 x 200 $\mu m^2$ and a sample number of 16-32 are used. Spectra of the aggregates are registered to the same spot in the bright-field. Post-processing involves filtering with a linear, third-order Savitzky-Golay filter, restricting the wavenumber range of the spectrum, and background-subtraction by fitting with a second-order polynomial. The peak position of the MIR resonance is extracted by fitting a Voigt function to the spectrum. The amplitude, peak position and phase of the Fano resonances are extracted using a physical fit (SI Figure S21).

**Raman microscopy.** SERS measurements were collected on a Renishaw InViva Raman confocal microscope, using a 20x objective (NA = 0.4) and 1200 lines $mm^{-1}$ grating. To collect the lower wavenumber spectral regions, the 785 nm excitation laser was used in extended scan mode, with 1 s integration time and 0.5 % laser power (2.2 mW at the sample). All measurements were taken at room temperature and the spectra were calibrated with respect to Si. Spectra were analyzed using Python, and any background removed by iteratively fitting a polynomial to the base of the peaks.

**Characterization.** Scanning electron microscope (SEM) imaging of MLaggs deposited on graphene and ZnSe uses a FEI Philips Dualbeam Quanta 3D SEM (dwell 3–10 µs, HV 2 kV, current 50 pA, and ≈3.0 mm working distance).

**Density Functional Theory (DFT) calculations:** DFT calculations on ferrocyanide were performed using the B3LYP[40,41] hybrid generalized gradient approximation exchange-correlation functional, augmented with Grimme's D3 dispersion correction and Becke-Johnson damping (GD3BJ).[42] The Def2SVP basis set was employed for all atoms.[43] A DFT scaling factor of 0.985 was applied to all calculated vibrational frequencies and the peaks broadened with a Gaussian fit. All DFT calculations were implemented using an ultrafine integration grid in Gaussian 09 Rev. E.[44] Due to the local plasmonic near field orientation being vertical to the gold surface, the IR and Raman modes for the adsorbed species will have different intensities relative to an isotropic system. These were calculated using the methodology in [45] from the IR dipoles and Raman tensors extracted from the DFT simulations. DFT calculations on $Au_2O_3$ were carried out using the Vienna Ab initio Simulation Package (VASP)[46–48] with core-valence electron interactions described via the projector augmented wave (PAW) method.[49] The Au and O pseudopotentials were taken from the PAW_PBE_64 library. All calculations employed the PBEsol exchange-correlation functional[50] with a Hubbard U parameter of 0.702 eV applied to the Au d orbitals[51] to account for localized electron correlation. A plane wave cutoff energy of 600 eV was used (increased to 780 eV for variable-cell calculations), with a k-point sampling density of 0.32 $Å^{-1}$. Atomic positions were optimized until residual forces fell below $5 \times 10^{-4}$ eV $Å^{-1}$. Force constants and IR intensities were computed using phonopy[52,53] and the phonopy-spectroscopy package[54] for pre- and post-processing of the lattice dynamics calculations.

**COMSOL simulations.** To calculate the field enhancement and the effective refractive index of the metamaterial we conduct frequency domain simulations. The simulation domain is a hexagonal unit cell, with Floquet periodic boundary conditions applied on the sides and a perfectly matched layer applied at the top and bottom. An incoming port is added at the bottom to launch the excitation and obtain the complex reflection coefficient, and an outgoing port is added at the top to obtain the complex transmission coefficient. The ports are set to be half a wavelength away from the metamaterial layer. The metamaterial layer is set in air, and it consists of the plasmonic gold resonator with diameter $D$, embedded in a slab of refractive index $n$ = 1.5, thickness $d$. The effective refractive index of the metamaterial is found from the complex reflection and transmission coefficients by standard inversion technique.[55] The field enhancement is extracted along the center line of a boundary of the unit cell. To calculate the decay of the field inside the homogeneous slab we set its thickness to $d$ = 677 nm (corresponding to 8 layers of tightly packed nanoparticles) and based on the retrieved effective parameters at $\lambda$ = 785 nm we set its refractive index to $n$ = 4.7 + 3.7i.

ASSOCIATED CONTENT

**Supporting Information**.

Supplementary Figs. 1–22, Supplementary Note 1

AUTHOR INFORMATION

## Corresponding Author

* Prof Jeremy J Baumberg, jjb12@cam.ac.uk

## Contributions

N.S., T.J., S.M.S.T, R.A. and J.J.B conceived and designed the experiment. N.S., T.J. and S.M.S.T performed the fabrication and spectroscopic experiments with input from R.A. and Y.R., with C.T. and S.Z. aiding the characterization. N.S. and T.J. analyzed the data with input from J.J.B., R.A., I.Q.L. Simulations and analytical modelling were carried out by Z.S., N.S., T.J. and C.T.. All authors contributed to writing and editing the manuscript.

## Notes

The authors J.J.B., S.M.S.-T. and R.A. declare the following competing interests: filed patents, Surface-enhanced spectroscopy substrates, UK 2304765.7, 30/3/2023; Detection method and in-flow Detection Method, UK Patent Application No: 2418909.4, 20/12/2024.

## ACKNOWLEDGMENT

The authors acknowledge financial support from the European Research Council (ERC) under Horizon 2020 research and innovation programme PICOFORCE (Grant Agreement No. 883703), from the UKRI

(UKRI3173, UKRI129885R) and from the EPSRC (Cambridge NanoDTC EP/L015978/1, EP/X037770/1, EP/Y008162/1, EP/Y036379/1, UKRI1255). N.S. acknowledges support from EPSRC Grant EP/L015889/1 for the EPSRC Centre for Doctoral Training in Sensor Technologies and Applications, and from AstraZeneca (MedImmune Ltd). S.M.S.-T. is supported by the Harding Distinguished Scholarship Programme and Magdalene College Cambridge. C.T. is supported by a Gates Cambridge fellowship (OPP1144). R.A. acknowledges support from the Winton Programme for the Physics of Sustainability and from St. John's College Cambridge. A.D. wishes to acknowledge support from the Royal Society University Research Fellowship URF/R1/180097 and URF/R/231024, Royal Society Research Fellows Enhancement Award RGF /EA/181038, funding from EPSRC grants EP/Y008774/1 and EP/X012689/1, and for the CDT in Topological Design EP/S02297X/1. Q.L. acknowledges support from the Dutch Research Council (NWO OCENW.XS24.2.142, 24.4.328) and PhotonDelta (NGF.1708.25.029).